\documentclass[aps,amsmath]{revtex4}
\usepackage{t1enc}
\usepackage[latin1,cp1251]{inputenc}
\usepackage[english]{babel}
\usepackage{graphics,epsfig}
\usepackage{natbib}
\usepackage{bm}

\begin{document}

\title{Neutron - antineutron oscillations in a trap revisited}
\author{B.~O.~Kerbikov}\email{borisk@heron.itep.ru}
\author{A.~E.~Kudryavtsev}\email{kudryavt@heron.itep.ru}
\author{V.~A.~Lensky}\email{lensky@vxitep.itep.ru}

\address{State Research Center Institute of Theoretical and Experimental Physics,
Bol'shaya Cheremushkinskaya 25, 117218 Moscow, Russia}

\begin{abstract}
We have reexamined the problem of $n-\bar n$ oscillations for ultra-cold neutrons (UCN) confined
within a trap. It is shown that for up to $10^3$ collisions with the walls the process can be described in
terms of wave packets. The exact equation is derived describing the time-evolution of the $\bar n$-component
for an arbitrary number of collisions with the trap walls.
The $\bar n$-component grows linearly with time with enhancement factor
depending on the reflection properties of the walls.
\end{abstract}



\maketitle

\section{Introduction}

For quite a long time physics beyond the Standart Model continues to be an intriguing subject. Several reactions which may serve
as signatures for the new physics have been discussed. One of the most elegant proposal is to look for $n-\bar n$ oscillations \cite{Kuzmin}
(see also \cite{Marshak}). There are three possible experimental settings aimed at the observation of this process. The first one is to establish
a limit on nuclear instability since $\bar n$ produced inside a nucleus will blow it up. The second one is to use a neutron beam from a reactor.
This beam propagates a long distance to the target in which possible $\bar n$ component would annihilate and thus be detected. 
The third option which we discuss in the present paper is to use ultra-cold neutrons (UCN) confined in a trap. The main question
is to what extent the generation of the $\bar n$ component is reduced by the interaction with the trap walls. This subject was addressed
by several authors \cite{Chetyrkin, Marsh, Baldo, Golub, Yoshiki, Kerbikov}. A thorough investigation of the problem is in our opinion still lacking.

First of all the clear formulation of the problem of $n-\bar n$ oscillations in a cavity has been hither to missing. Two different approaches were employed
without presenting sound arguments in favor of their applicability and without tracing connections between them.

In the first approach \cite{Marsh, Baldo} $n-\bar n$ oscillations are considered in the basis of the discrete eigenstates of the trap potential taking
into account the splitting between $n$ and $\bar n$ levels and $\bar n$ annihilation. The density of the trap eigenstates which is proportional to
the macroscopic trap volume is huge and the states cluster together extremely thickly. However these arguments are not enough to discard
the discrete states approach since $n-\bar n$ mixing parameter is much smaller than the distance between adjacent levels --- see below.
The true reason due to which the discrete eigenstates approach is of little physical relevance is the following. The spectrum of the neutrons provided to the
trap by the source is continuous and certain time is needed for the rearrangement of the initial wave function into standing waves corresponding
to the trap eigenstates. As will be shown below this time interval is of the order of $\beta$-decay time so that the standing waves regime being
interesting by itself can hardly be reached in the real physical situation.

\begin{figure}[h]
\epsfbox{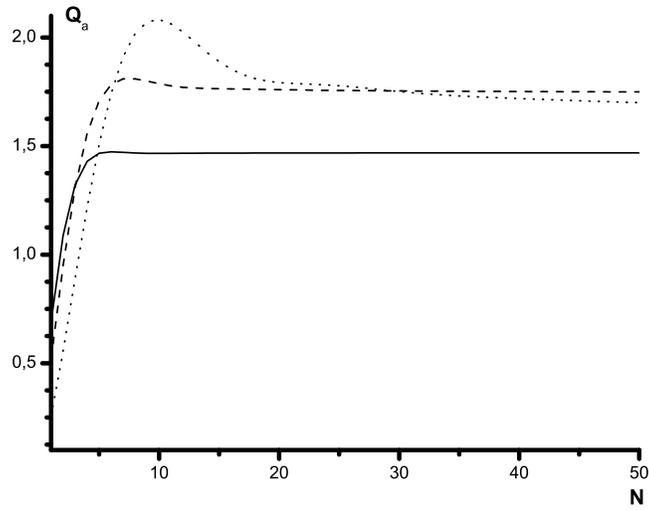}
\caption{Plot of $Q_a(N)=(\epsilon^2\tau_L t)^{-1}P_a(N)$ dependence versus $N$. Solid line corresponds to $^{12}\mathrm C$ (graphite),
dashed one - to $^{12}\mathrm C$ (diamond), dotted one - to $\mathrm{Cu}$.}
\label{Fig1}
\end{figure}

The second approach \cite{Chetyrkin, Golub, Yoshiki} treats the neutrons and antineutrons inside a trap as freely moving particles which
undergo reflections from the trap walls. Collisions with the walls result in a reduction of the $\bar n$ component compared to the case of the
free-space evolution. This suppression is due to two factors. The first one is the annihilation inside the walls. The second one is the phase
decoherence of $n$ and $\bar n$ components induced by the difference of the walls potentials acting on $n$ and $\bar n$. Reflections
of antineutrons from the trap walls were for the first time considered in \cite{Chetyrkin}. The purpose of that paper was to investigate
the principal possibility to observe $n-\bar n$ oscillations in a trap and the authors estimated reflection coefficient for antineutrons
without paying attention to the decoherence phenomena. Only single collision with the trap wall was considered in~\cite{Chetyrkin}.
A comprehensive study of $n-\bar n$ oscillations in a trap was presented in~\cite{Golub,Yoshiki}. Decoherence and multiple reflections
were included as well as the influence of the gravitational and magnetic fields. The approximate equation for the annihilation probability after $N$ collisions
was obtained in~\cite{Yoshiki} (equation (3.8)) which coincides with the exact formula (59) of the present paper when $N\gg 1$. As we show below
(see Fig.~\ref{Fig1}) the $N$-independent asymptotic regime settles at $N\agt 10$.

The derivation of the exact equation for the annihilation probability for arbitrary number of collisions is in no way the only purpose of the present work.
We already mentioned the problem of the relation between the eigenvalue and the wave packet approaches. Within the wave packet approach
some basic notions such as the time between successive collisions and the collision time itself can be defined in a clear and rigorous way.
Another question within the wave-packet formalism is the independence of the reflection coefficient on the width of the wave packet and the applicability
of the stationary formalism to calculate reflections from the trap walls. These and some other principal points are considered for the first time
in detail in the present paper.

Let us also mention that an alternative approach to the evaluation of the reflection coefficients for $n$ and $\bar n$ was outlined in~\cite{Kerbikov}.
Calculations presented there are based on the time-dependent Hamilton formalism for the interaction of $n$ and $\bar n$ with the trap walls.
This subject remains outside the scope
of the present paper.

The paper is organized as follows. In Section II we remind the basic equations describing $n-\bar n$ oscillations in free space.
Section III is devoted to the optical potential approach to the interaction of $n$ and $\bar n$ with the trap walls. In Section IV we analyze
the two formalisms proposed to treat $n-\bar n$ oscillations in the cavity, namely box eigenstates and wave packets. In Section V the
reflection from the trap walls is considered. Section VI contains the main result of this work which is the time dependence of the 
$\bar n$~component production probability. In Section VII conclusions are formulated and problems to be solved outlined.

\section{Oscillations in Free Space}

We start by reminding the basic equations describing $n-\bar n$ oscillations in free space. The phenomenological Hamiltonian is a $2\times 2$
matrix in the basis of the two-component $n-\bar n$ wave function (we set $\hbar = 1$)

\begin{equation}
H_{jl}=(H_j-i\frac{\Gamma_\beta}{2})\delta_{jl}+\epsilon (\sigma_x)_{jl},
\end{equation}
where $j,l=n,\bar n$; $H_j=k^2/2m-\mu_j B$, $\mu_j$ is the magnetic moment, $B$ is the external (e.g.\ the Earth) magnetic field, $\Gamma_\beta$
is the $\beta$-decay width, $\epsilon$ is the $n-\bar n$ mixing parameter (see below), $\sigma_x$ is the Pauli matrix. Assuming the $n$ and $\bar n$
wave functions to be plane waves we write the two-component wave function of the $n-\bar n$ system as

\begin{equation}
\hat\Psi(x,t)=
\begin{pmatrix}
\psi_n(t)\cr \psi_{\bar n}(t)
\end{pmatrix}
e^{ikx}.
\end{equation}
Then the evolution of the time-dependent part of $\hat \Psi(x,t)$ is described by the equation

\begin{equation}
i\frac{\partial}{\partial t}
\begin{pmatrix}
\psi_n(t)\cr\psi_{\bar n}(t)
\end{pmatrix}
=
\begin{pmatrix}
E_n-i\frac{\Gamma_\beta}{2} & \epsilon\cr
\epsilon & E_{\bar n}-i\frac{\Gamma_\beta}{2}
\end{pmatrix}
\begin{pmatrix}
\psi_n(t)\cr\psi_{\bar n}(t)
\end{pmatrix}
\end{equation}
The difference between $E_n$ and $E_{\bar n}$ due to the Earth magnetic field is

\begin{equation}
\omega=E_{\bar n}-E_n=2|\mu_n|B\approx 6\times 10^{-12} \ \mathrm{eV}.
\end{equation}
Diagonalizing the matrix in (3) we find $\psi_n(t)$ and $\psi_{\bar n}(t)$ in terms of their values at $t=0$:

\begin{equation}
\psi_n(t)=\left(\psi_n(0)(\cos \nu t +\frac{i\omega}{2\nu}\sin \nu t)-\psi_{\bar n}(0)\frac{i\epsilon}{\nu}\sin \nu t\right)
\exp[-\frac{1}{2}(i\Omega+\Gamma_\beta)t]
\end{equation}
\begin{equation}
\psi_{\bar n}(t)=\left(-\psi_n(0)\frac{i\epsilon}{\nu}\sin \nu t+\psi_{\bar n}(0)(\cos \nu t -\frac{i\omega}{2\nu}\sin \nu t)\right)
\exp[-\frac{1}{2}(i\Omega+\Gamma_\beta)t],
\end{equation}
where $\Omega = E_n+E_{\bar n}$, $\nu=(\omega^2/4 + \epsilon^2)^{1/2}$, $\omega=E_{\bar n}-E_n$. In particular if $\psi_n(0)=1, \psi_{\bar n}(0)=0$,
one has

\begin{equation}
|\psi_{\bar n}(t)|^2=\frac{4\epsilon^2}{\omega^2+4\epsilon^2}\exp(-\Gamma_\beta t)\sin^2(\frac{1}{2}\sqrt{\omega^2+4\epsilon^2}t).
\end{equation}
The use of this equation to test fundamental symmetries is discussed in \cite{Abov}.

Without magnetic field, i.e.\ for $\omega=0$, and for $t\ll\epsilon^{-1}$ equation (7) yields

\begin{equation}
|\psi_{\bar n}(t)|^2\approx\epsilon^2t^2\exp(-\Gamma_\beta t).
\end{equation}
This law (for $t\ll\Gamma^{-1}_\beta$) has been employed to establish the lower limit on the oscillation time $\tau=\epsilon^{-1}$.
According to the ILL-Grenoble experiment \cite{Baldoetal}

\begin{equation}
\tau>0.86\times 10^8\ \mathrm{s}.
\end{equation}
The corresponding value of the mixing parameter is $\epsilon\approx 10^{-23}\ \mathrm{eV}$.
This number will be used in obtaining numerical results presented below.

The Earth magnetic field leads to a strong suppression of $n-\bar n$ oscillations. With the value of $\omega$ given by (4)
equation (7) leads to the following result:

\begin{equation}
|\psi_{\bar n}(t)|^2\approx\frac{4\epsilon^2}{\omega^2}\exp(-\Gamma_\beta t)\sin^2t/\tau_B\approx10^{-23}\sin^2t/\tau_B,
\end{equation}
where $\tau_B=(|\mu_n|B)^{-1}\approx2\times 10^{-4}\ \mathrm{s}$. In what follows we shall assume that magnetic field is screened.

For $\omega=0$ but for arbitrary initial conditions equations (5--6) take the form

\begin{equation}
\psi_n(t)=\left(\psi_n(0)\cos \epsilon t -i \psi_{\bar n}(0)\sin \epsilon t\right)\exp[-(iE+\frac{\Gamma_\beta}{2})t]
\end{equation}
\begin{equation}
\psi_{\bar n}(t)=\left(-i\psi_n(0)\sin \epsilon t+\psi_{\bar n}(0)\cos \epsilon t\right)\exp[-(iE+\frac{\Gamma_\beta}{2})t],
\end{equation}
where $E=E_n=E_{\bar n}$.

\section{Optical Potential Model for the Trap Wall}

We remind that neutrons with energy $E<10^{-7}\mathrm{eV}$ are called ultra-cold.  An excellent review of UCN physics was given by I.~M.~Frank
\cite{IFrank} (see also \cite{Ignatovich}). 

A useful relation connecting the neutron velocity $v$ in cm/s and $E$ in eV reads

\begin{equation}
v(\mathrm{cm/s})=10^2[10^9 E(\mathrm{eV})/5.22]^{1/2}.
\end{equation}
For $E=10^{-7}\ \mathrm{eV}$ the velocity is $v\approx 4.4\times 10^2\ \mathrm{cm/s}$.

A less formal definition of UCN involves a notion of the real part of the optical potential corresponding to the trap material (see below).
Neutrons with energies less than the height of this potential are called ultra-cold. The two definitions are essentially equivalent since
as we shall see for most materials the real part of the optical potential is of the order of $10^{-7}$ eV.

Our main interest concerns strongly absorptive interaction of the $\bar n$ component with the trap walls. Therefore wery weak absorption
of UCN on the walls \cite{IFrank, Ignatovich} will be ignored. Due to complete reflection from the trap walls UCN can be stored for about
$10^3$ s ($\beta$-decay time) as was first pointed out by Ya.~B.~Zeldovich \cite{Zeld}.

To be concrete we consider UCN with $E=0.8\times 10^{-7}$ eV which corresponds to $v=3.9\times 10^2$ cm/s (see (13)), $k=12.3$ eV and 
de Broglie wave length $\lambda\approx 10^{-5}$ cm. In the next Section we shall describe UCN in terms of the wave packets and hence 
the above values should be attributed to the center of the packet.

Interaction of $n$ and $\bar n$ with the trap walls will be treated in terms of energy-independent optical potential. The validity of this 
approach to UCN has been justified in a number of papers, see e.g.\ \cite{IFrank, Ignatovich, Lax}. There is still an open question concerning
the discrepancy between theoretical prediction and experimental data on UCN absorption. Interesting by itself this problem is outside 
the scope of our work since as already mentioned absorption of neutrons may be ignored in $n-\bar n$ oscillation process.
The low-energy optical potential reads

\begin{equation}
U_{jA}=\frac{2\pi}{m}Na_{jA},
\end{equation}
where $j=n, \bar n$; $m$ is the neutron mass, $N$ is the number of nuclei in a unit volume, $a_{jA}$ is the $j-A$ scattering length which
is real for $n$ and complex for $\bar n$. For neutrons the scattering lengths $a_{nA}$ are accurately known for various materials \cite{Ignatovich}.
For antineutrons the situation is different. Experimental data on low-energy $\bar n-A$ interaction are absent. Only some indirect
information may be gained from level shifts in antiprotonic atoms. Therefore the values of $a_{\bar n A}$ used in
\cite{Chetyrkin, Golub, Kerbikov, Hufner} as an input in the problem of $n-\bar n$ oscillations are similar but not the same. We consider as most
reliable the set of $a_{\bar n A}$ calculated in \cite{Kondratyuk} within the framework of internuclear cascade model. Even this particular
model leads to several solutions. Therefore the one we have chosen for $^{12}\mathrm C$ (graphite and diamond) may be called ''motivated''
by Ref.~\cite{Kondratyuk}. To get a feeling on the dependence upon the material of the walls and to compare our results with that of
\cite{Chetyrkin} we also made calculations for $\mathrm{Cu}$. Scattering lengths for $\mathrm{Cu}$ are not presented in
\cite{Kondratyuk} and we used a solution proposed in \cite{Chetyrkin}. Thus our calculations were performed with the following set of $\bar n-A$
scattering lengths:

\begin{equation}
a_{\bar n\mathrm C}=(3-i1) \ \mathrm{fm},\,\, a_{\bar n\mathrm{Cu}}=(5-i0.5)\ \mathrm{fm}.
\end{equation}
The scattering lengths for neutrons are \cite{Ignatovich}:
\begin{equation}
a_{n\mathrm C}=6.65 \ \mathrm{fm},\,\, a_{n\mathrm{Cu}}=7.6\ \mathrm{fm}.
\end{equation}
The concentrations of atoms $N$ entering into (14) are: $N_{\mathrm{C(graphite)}}=1.13\times 10^{-16}\ \mathrm{fm}^{-3}$, $
N_{\mathrm{C(diamond)}}=1.63\times 10^{-16}\ \mathrm{fm}^{-3},\ N_{\mathrm{Cu}}=0.84\times 10^{-16}\ \mathrm{fm}^{-3}$. Then according to (14)
the optical potentials read

\begin{equation}
\begin{split}
U_{n\mathrm{C(gr)}}=1.95\times 10^{-7}\ \mathrm{eV},\\
U_{n\mathrm{C(diam)}}=2.8\times 10^{-7}\ \mathrm{eV},\\
U_{n\mathrm{Cu}}=1.66\times 10^{-7}\ \mathrm{eV};
\end{split}
\end{equation}
\begin{equation}
\begin{split}
U_{\bar n\mathrm{C(gr)}}=(0.9-i0.3)\times 10^{-7}\ \mathrm{eV},\\
 U_{\bar n\mathrm{C(diam)}}=(1.3-i0.4)\times 10^{-7}\ \mathrm{eV},\\ 
U_{\bar n\mathrm{Cu}}=(2-i0.2)\times 10^{-7}\ \mathrm{eV}.
\end{split}
\end{equation}
In the present work we consider particles ($n$ and $\bar n$) with energies below the potential barrier formed by the real part of the potential.
For $\bar n$ and $^{12}\mathrm C$ the limiting velocity is $v=4.15\times 10^2$ cm/s.

\section{Wave Packet versus Standing Waves}

It is convenient to use for the optical potentials (17) and (18) the shorthand notation

\begin{equation}
U_j=V_j-iW_j\delta_{j\bar n},
\end{equation}
where $j=n,\bar n$ while the wall material is not indicated explicitly. We consider the following model for the trap in which $n-\bar n$ 
oscillations may be possibly observed. Imagine the two walls of the type (19) separated by a distance $L\sim 10^2$ cm, i.e.\ the one-dimensional 
potential well of the form

\begin{equation}
U_j(x)=\left\{\theta(-x-L)+\theta(x)\right\}\left\{V_j-iW_j\delta_{j\bar n}\right\},
\end{equation}
with $\theta(x)$ being a step function. Our goal is to follow the time evolution of the $\bar n$ component in such a trap assuming that the initial
state is a pure $n$ one.

The first question to be answered is how to describe the wave function of the system. Two different approaches seem to be feasible and both
were discussed in the literature \cite{Marsh, Golub, Kerbikov}. The first one is to consider oscillations occuring in the wave packet and to
investigate to what extent reflections from the walls distort the picture as compared to the free-space regime. The second one is to consider 
eigenvalue problem in the potential well (20), to find energy levels for $n$ and $\bar n$, and to consider oscillations in this basis. Due to different
interaction with the walls the levels of  $n$ and $\bar n$ are splitted and $\bar n$ levels acquire annihilation widths.

At first glance this approach might seem inadequate since in a trap with $L\sim 10^2$ cm the density of states is very high, the characteristic
quantum number corresponding to the UCN energy is very large, the splitting $\delta E$ between adjacent $n$-levels (or between the levels of
$n$ and $\bar n$ spectra) is extremely small. The values of all these quantities will be given below and we shall see that $\delta E<10^{-14}$ eV.
However, this approach can not be discarded without further analysis since the $n-\bar n$ mixing parameter $\epsilon\approx 10^{-23}$ eV is
much smaller than $\delta E$.

To understand the relation between the two approaches note that the initial conditions correspond to a beam of UCN provided by a source.
The momentum spectrum of UCN depends on the concrete experimental conditions. In order to stay on general grounds and at the same time
to simplify the problem we shall assume that the UCN beam entering the trap has a form of a Gaussian wave packet. Suppose that at $t=0$
the center of the wave packet is at $x=x_0$ so that

\begin{equation}
\psi_k(x, t=0)=(\pi a^2)^{-1/4}\exp\left(-\frac{(x-x_0)^2}{2a^2}+ikx\right),
\end{equation}
where $a$ is the width of the wave packet in coordinate space. The normalization of the wave function (21) corresponds to one particle over
the whole one-dimensional space:

\begin{equation}
\int\limits_{-\infty}^{+\infty}dx|\psi_k(x,t=0)|^2=1.
\end{equation}

For $E=0.8\times 10^{-7}$ eV and the beam resolution $\Delta E/E=10^{-3}$ one has 

\begin{equation}
k=12.3\  \mathrm{eV}, \ \ a=3.2\times 10^{-3}\  \mathrm{cm}.
\end{equation}
The width of the wave packet (21) increases with time according to

\begin{equation}
a^{\prime}=a\left[1+\left(\frac{t}{ma^2}\right)^2\right]^{1/2}\approx t/ma
\end{equation}
and for $t\sim 10^3$ s becomes comparable with the trap size $L$. In order that the wave hitting the wall and the reflected one to be clearly
resolved the condititon $a^{\prime}/v\ll\tau_L$, or $a^\prime\ll L$ has to be satisfied, where $\tau_L\sim 1$ s is the time between two consecutive
collisions with the trap walls. Reflection of the wave packet from the walls is considered in detail in the next Section.
Here we will show that $t\sim 10^3$ s is the characteristic time needed for the rearrangement of the initial wave packet into the 
stationary states of the trapping box.

Consider the eigenvalue problem for the potential well (20). Parameters of the potential (20) for
neutrons are $V_n\approx 2\times 10^{-7}\ \mathrm{eV},\  L\approx10^2\ \mathrm{cm}$. The number of levels is

\begin{equation}
M\approx\frac{L\sqrt{2mV}}{\pi}\approx 10^8/\pi.
\end{equation}
According to (23) the center of the wave packet (21) has a momentum $k=12.3$ eV which corresponds to a state with a number of nodes 
$j\approx 2\times 10^7$ and $k_j L\approx 6\times 10^7\gg 1$. Positions of such highly excited levels in a finite depth potential are indistinguishable
from the spectrum in a potential box with infinite walls. Thus

\begin{equation}
\varphi_j(x)\approx\sqrt{\frac{2}{L}}\sin\omega_j x,\ \ \omega_j=\frac{\pi j}{L}.
\end{equation}
The wave functions (26) describe semi-classical states with $j\gg 1$ in a potential well with sharp edges.
The ''frequency'' $\omega_j$ is very high as compared to the width of the wave packet in momentum space, $\omega_j\approx 6\times 10^5\ 
\mathrm{cm}^{-1}\gg\nu=1/(\sqrt{2}a)\approx 2\times 10^{2}\ \mathrm{cm}^{-1}$. This means that the wave packet spans over a large number of levels.
To determine this number note that the distance between adjacent levels around the center of the wave packet is $\delta E=E_{j+1}-E_j\sim 10^{-14}$ eV.
The highly excited levels within the energy band $\Delta E=10^{-3}E\sim 10^{-10}$ eV corresponding to the wave packet (21) are to a high accuracy
equidistant as it should be in a semi-classical regime. The number of states within $\Delta E$ is $\Delta j=\Delta E/\delta E\sim 10^4$ and their
density in momentum space is 

\begin{equation}
\rho(\omega)=a\Delta j\simeq L/\pi\sim 10^6\ \mathrm{eV}^{-1}.
\end{equation}

Now we can answer the question formulated at the beginning of this Section, namely whether one should describe $n-\bar n$ oscillations in the
trap in terms of the wave packet or in terms of the stationary eigenfunctions. At $t=0$ the wave function has the form of the wave packet (21) provided by the UCN source.
Due to collisions with the trap walls transitions from the initial state (21) into discrete (or quasi-discrete for $\bar n$) eigenstates (26) take place.

The time-evolution of the initial wave function (21) proceeds according to

\begin{equation}
\psi(x,t)=\int dx^\prime G(x,t;x^\prime,0)\psi_k(x^\prime,0),
\end{equation}
where $G(x,t;x^\prime,0)$ is the time-dependent Green's function for the potential well (20).
Making use of the spectral representation for $G$, one can write

\begin{equation}
\psi(x,t)=\sum_je^{-iE_jt}\varphi_j(x)\int dx^\prime\varphi^*_j(x^\prime)\psi_k(x^\prime,0).
\end{equation}

In semi-classical approximation the distance between the adjacent levels is $\delta E=\pi/\tau_L$. Therefore one may think that at $t=\tau_L$,
i.e.\ already at the first collision the neighboring terms in (29) would cancel each other. However this is not the case. Indeed,
\begin{equation}
\varphi_{j+1}(x)e^{-iE_{j+1} t}+\varphi_j(x)e^{-iE_j t}
=\frac{e^{-iE_j t}}{i\sqrt{2L}}\left[e^{i\omega_j x}(1+e^{i\frac{\pi}{L}(x-vt)})
-e^{-i\omega_j x}(1+e^{-i\frac{\pi}{L}(x+vt)})\right].\nonumber
\end{equation}
Therefore at $x=\pm vt$ there is a constructive interference either in the first or in the second terms correspondingly. This holds true
with the account of the whole sum of terms in (29) and hence one can pass from summation to integration in (29). 
The wave function overlap entering into (29) can be easily evaluated provided the center of the wave packet $x_0$ is
not within the bandwidth distance $a^\prime$ from the trap walls. The overlap is given by the following integral:

\begin{equation}
\begin{split}
\int dx^\prime\varphi^*_j(x^\prime)\psi_k(x^\prime,0)
\simeq\frac{i}{(2\sqrt{\pi}La)^{1/2}}\int\limits^0_{-L}dx^\prime\exp\left(-\frac{(x^\prime-x_0)^2}{2a^2}+i(k-\omega_j)x^\prime\right)\\
=\frac{i}{(2\sqrt{\pi}La)^{1/2}}\int\limits_0^Ldx^{\prime\prime}\exp\left(-\frac{(x^{\prime\prime}+x_0)^2}{2a^2}-i(k-\omega_j)x^{\prime\prime}\right).
\end{split}
\end{equation}
At this step we have omitted the exponent with high frequency $(k+\omega_j)$. Next take $(x^{\prime\prime}+x_0)/(\sqrt{2}a)$ as
a new variable and resort to the assumption $|x_0|\gg a,\ L-|x_0|\gg a$ (we remind that $x_0$ is negative since $-L<x<0$).
The result reads

\begin{equation}
\int dx^\prime\varphi^*_j(x^\prime)\psi_k(x^\prime,0)
\simeq i(\frac{\sqrt{\pi}a}{L})^{1/2}\exp\left(-\frac{a^2}{2}(k-\omega_j)^2+i(k-\omega_j)x_0\right).
\end{equation}

Corrections to (31) are of order of $a/L$. Now comes frequency summation in (29). This summation can be substituted by integration
over $\omega$ since the density of semi-classical states $\rho(\omega)$ is very high. In this way we arrive at

\begin{equation}
\psi(x,t)=\frac{1}{[\sqrt{\pi}a(1+i\frac{t}{ma^2})]^{1/2}}\Biggl(\exp[-\frac{\alpha(x,t)}{2a^2(1+\frac{t^2}{m^2a^4})}]
+\exp[-\frac{\alpha(-x,t)}{2a^2(1+\frac{t^2}{m^2a^4})}]\Biggl),
\end{equation}
\begin{equation}
\alpha(x,t)=(x-x_0-v_0t)^2-it\frac{(x-x_0)^2}{ma^2}-2ik_0a^2(x-x_0)+i\frac{k_0^2 a^2}{m}t+2ik_0x_0(a^2+\frac{t^2}{m^2a^2}).
\end{equation}

The second term in (32) describes the reflected wave packet --- see the next Section. According to (21), (28) and (32) all that
happens to the wave packet in the trap is broadening and reflections. This is true at least during some initial period of
its life history. How long does this period last? The answer to this question may be obtained estimating the accuracy
of performing frequency integration instead of summation over discrete states in (29).

To estimate the time scale for the rearrangement of the initial wave packet (21) into the trap standing waves (26), 
it is convenient to introduce the difference $\delta\psi(x,t)=\psi_{sum}(x,t)-\psi_{int}(x,t)$ between the ''exact''
wave function (29) and the approximate integral representation (32). As soon as

\begin{equation}
\delta w(t)=\int dx\left(|\psi_{sum}|^2-|\psi_{int}|^2\right)=2\int dx \Re(\psi_{int}\delta\psi)\ll 1,
\end{equation}
one can consider oscillations as proceeding in the wave packet basis. The following estimate holds

\begin{equation}
\begin{split}
\delta\psi(x,t)=\sum_n f(\omega_n)-\int d\omega\rho(\omega)f(\omega)
=-\sum_n\int\limits^{\omega_{n+1}}_{\omega_n}d\omega\rho(\omega)(f(\omega)-f(\omega_n))\\
\simeq-\sum_n\int\limits^{\omega_{n+1}}_{\omega_n}d\omega\rho(\omega)f^\prime(\omega_n)(\omega-\omega_n)
=-\frac{1}{2}\sum_nf^\prime(\omega_n)(\omega_{n+1}-\omega_n);
\end{split}
\end{equation}

here
\begin{equation}
f(\omega)=\sqrt{\frac{\sqrt{\pi}a}{2L^2}}
\exp\left(-\frac{a^2}{2}(k_0-\omega)^2-i\frac{\omega^2}{2m}t+i\omega (x-x_0)+ikx_0\right).
\end{equation}

From (36) one gets
\begin{equation}
\begin{split}
f^\prime(\omega)=g(\omega)f(\omega),\\
g(\omega)=i(x-x_0-vt)-(k_0-\omega)a^2.
\end{split}
\end{equation}
Since $f(\omega)$ is a narrow Gaussian peak, we can substiute $g(\omega)$ by $g(k_0)$, and then (35)
results in 

\begin{equation}
\delta\psi(x,t)\simeq\frac{\pi}{2L}(x-x_0-v_0t)\psi_{int}(x,t).
\end{equation}

From (34) and (38), we get

\begin{equation}
\delta w\simeq\frac{\pi}{2L}\int\limits^{+\infty}_{-\infty} dx|x-x_0-v_0t||\psi_{int}(x,t)|^2
\propto\frac{a^\prime}{L}\sim\frac{t}{maL}\sim \frac{t}{10^3\thinspace\mathrm{s}},
\end{equation}
where $a^\prime$ is given by (23).

Roughly speaking, the time $t\sim 10^3$ s needed for the neutron wave function to rearrange into the trap eigenstate
is comparable to the neutron life-time and neutron would rather ''die'' than adjust to the new boundary conditions.
Therefore in what follows the wave packet formalism will be used. Some additional subtleties arising from the quantization
of the levels in the trapping box will be discussed in Section VII.

\section{Reflection from the Trap Walls}

Consider again a one-dimensional trap (20). Let the particle moving from $x=-\infty$ enter the trap at $t=0$ through the window at $x=-L$. At $t=\tau_L$ it will reach the
wall at $x=0$, the $n$ component will be reflected from the wall while the $\bar n$ component will be partly reflected and partly
absorbed. The wave packet describing the interaction with the wall has the form
 
\begin{equation}
\psi(x,t)=\pi^{-3/4}\sqrt{\frac{a}{2}}\int\limits^{+\infty}_{-\infty}dk\psi_j(k,x)\exp\left(-\frac{a^2}{2}(k-k_0)^2+iL(k-k_0)-i\frac{t}{2m}k^2\right),
\end{equation}
where $j=n,\bar n$ and

\begin{equation}
\psi_j(k,x)=e^{ikx}+R(k)e^{-ikx}=e^{ikx}+\rho_j(k)e^{\phi_j(k)}e^{-ikx}.
\end{equation}
For the $n$ component $\rho_n(k)=1$ since we neglect very weak absorption of neutrons at the surface. The integral (40) with the first
term of (41) is trivial. In order to integrate the second term of (41) note that due to the Gaussian form-factor with $ak_0\sim 10^3\gg 1$
the dominant contribution to the integral (40) comes from the narrow interval of $k$ around $k_0$. Expanding $R_j(k)$ at $k-k_0$ and
keeping the leading term we get

\begin{equation}
\begin{split}
R_j(k)\simeq \rho_j(k_0)e^{i\phi_j(k_0)}[1+i\phi_j^\prime(k_0)(k-k_0)
+\delta_{j\bar n}\frac{\rho_j^\prime(k_0)}{\rho_j(k_0)}(k-k_0)]\\
\simeq\rho_j(k_0)e^{i\phi_j(k_0)+i\phi_j^\prime(k_0)(k-k_0)}.
\end{split}
\end{equation}
The validity of the last step for $\bar n$ will become clear from the explicit expression for $\rho_{\bar n}(k)$ and $\phi_{\bar n}(k)$
presented below.

Now integration in (40) can be easily performed with the result \cite{Galitsky}

\begin{equation}
\psi_j(x,t)=\frac{1}{[\sqrt{\pi}a(1+i\frac{t}{ma^2})]^{1/2}}\Biggl(\exp[-\frac{\alpha_{inc}(x,t)}{2a^2(1+\frac{t^2}{m^2a^4})}]+
R_j(k_0)\exp[-\frac{\alpha_{refl}(x,t)}{2a^2(1+\frac{t^2}{m^2a^4})}]\Biggl),
\end{equation}
\begin{equation}
\alpha_{inc}(x,t)=(x+L-v_0t)^2-it\frac{(x+L)^2}{ma^2}-2ik_0a^2(x+L)
+i\frac{k_0^2 a^2}{m}t+2ik_0L(a^2+\frac{t^2}{m^2a^2}),
\end{equation}
\begin{equation}
\alpha_{refl}(x,t)=\alpha_{inc}(-x+\phi^\prime,t)+2ik_0\phi^\prime(a^2+\frac{t^2}{m^2a^2}).
\end{equation}

From (43)--(45) we see that the essence of $R(k)$ in the wave packet formalism is the same as in the time independent approach.
Therefore imposing standard boundary conditions at $x=0$ we get the reflection coefficients

\begin{equation}
R_j(k)=\rho_j(k)e^{i\phi_j(k)}=\frac{k-i\kappa_j}{k+i\kappa_j},
\end{equation}
\begin{equation}
\begin{split}
\kappa_n=[2m(V_n-E)]^{1/2},\\
\kappa_{\bar n}=[2m(V_{\bar n}-iW_{\bar n}-E)]^{1/2}=\kappa^\prime_{\bar n}-i\kappa^{\prime\prime}_{\bar n},
\end{split}
\end{equation}
\begin{equation}
\tan \phi_n=\frac{-2k\kappa_n}{k^2-\kappa_n^2},
\ \tan \phi_{\bar n}=\frac{-2k\kappa^\prime_{\bar n}}{k^2-(\kappa_{\bar n}^\prime)^2-(\kappa_{\bar n}^{\prime\prime})^2},
\end{equation}
\begin{equation}
\rho_n=1,\ \rho_{\bar n}^2=1-\frac{4k\kappa^{\prime\prime}_{\bar n}}{(k+\kappa^{\prime\prime}_{\bar n})^2+(\kappa^\prime_{\bar n})^2}.
\end{equation}

In particular for $^{12}\mathrm{C}$ (graphite)
\begin{equation}
\rho=0.56,\ \theta\equiv\phi_{\bar n}-\phi_n=0.72.
\end{equation}

The first term in the right-hand side of (45) may be written as $[-x+L-v_0(t-\phi^\prime/v_0)]^2$. Hence the collision time or time-delay
is \cite{Galitsky, Goebel}

\begin{equation}
\tau_{j, \mathrm{coll}}=\phi^\prime_j(k_0)/v_0=\mbox{Re}\frac{2m}{k\kappa_j}.
\end{equation}
For neutrons, i.e.\ for real $\kappa_n$, equation (51) gives the well-known result $\tau_{n, \mathrm{coll}}=[E(V_n-E)]^{-1/2}$. This result
is in line with the naive estimate $\tau_{n, \mathrm{coll}}\sim l/v_0\sim 10^{-8}$ s \cite{Kerbikov}, where $l\alt\lambda$ is the penetration
depth.

For $^{12}\mathrm{C}$ (graphite) equation (51) yields

\begin{equation}
\tau_{n, \mathrm{coll}}=0.7\times 10^{-8}\ \mathrm{s},\ \tau_{\bar n, \mathrm{coll}}=1.1\times 10^{-8}\ \mathrm{s}.
\end{equation}

Equations (43)--(45) supplemented by the above inequality allow to follow the time evolution of the beam inside the trap.
Imagine an observer placed at a bandwidth distance from the wall, i.e.\ at $x=-a$. According to (43)--(45) such an observer
will conclude that at times $t\le\tau_L-\tau_a$ the incident wave (the first term in (43)) dominates, while at $t\ge \tau_L+\tau_a$
the reflected wave prevails. With this splitting of the time interval around $N\tau_L,\ N=1,2,\dots$ in mind we shall use the
notations $(N\tau_L-)$ and $(N\tau_L+)$ for the moments before and after the $N^\mathrm{th}$ collision. Thus, we can calculate 
$\bar n$ production rate since we have a rigorous definitions of the collision time and the time interval between the two
subsequent collisions.

\section{Annihilation Rate in a Trap}

Now we can inquire into the problem of the time-dependence of $\bar n$ production probability. In free space it is given by $|\psi_{\bar n}(t)|^2=
\epsilon^2 t^2$ (see (2)) while in a trap with complete annihilation or total loss of coherence at each collision it has a linear time dependence 
$|\psi_{\bar n}(t)|^2=\epsilon^2\tau_L t$~\cite{Kerbikov}.

To avoid cumbersome equations and in view of the fact that we consider time interval $t\ll\Gamma_\beta^{-1}$ we omit $\exp(-\Gamma_\beta t)$
factors. Production of $\bar n$ during the collision can be also neglected \cite{Kerbikov}. Difference in collision times (52) for $n$ and $\bar n$
may be ignored as well. In the previous Section we have seen that the interaction of the wave packet with the wall is described in terms of the
reflection coefficients (46).~\footnote{An alternative description using time-evolution operators was proposed in \cite{Kerbikov}.}

Assume that at $t=0$ a pure $n$ beam enters the trap at $x=-L$. After crossing the trap, i.e.\ at $t=(\tau_L-)$, the time-dependent parts of the wave
functions are given by (12) [we state this because although the Gaussian form factor in (43) also depends on time, the corresponding
terms in the time-dependent Shr\"odinger equation are of order of $1/ak_0$ as compared to the derivative of the exponent $\exp(-iEt)$;
also note that the form factors are the same for $n$ and $\bar n$ up to constant multiplier]:

\begin{equation}
\begin{split}
\psi_n(\tau_L-)=\cos(\epsilon \tau_L)\exp(-iE\tau_L),\\
\psi_{\bar n}(\tau_L-)=\sin(\epsilon \tau_L)\exp[-i(E\tau_L+\pi/2)].
\end{split}
\end{equation} 
After the first reflection at $t=(\tau_L+)$ we get

\begin{equation}
\begin{split}
\psi_n(\tau_L+)=\cos(\epsilon \tau_L)\exp[-i(E\tau_L-\phi_n)],\\
\psi_{\bar n}(\tau_L+)=\rho_{\bar n}\sin(\epsilon \tau_L)\exp[-i(E\tau_L-\phi_{\bar n}+\pi/2)].
\end{split}
\end{equation}
Evolution from $t=(\tau_L+)$ to $t=(2\tau_L-)$ again proceeds according to (12)

\begin{equation}
\begin{split}
\psi_{\bar n}=\frac{1}{2}\sin(2\epsilon\tau_L)\left(1+\rho e^{i\theta}\right)\exp[-i(2E\tau_L-\phi_{\bar n}+\pi/2)]\\
\simeq\epsilon\tau_L\left(1+\rho e^{i\theta}\right)\exp[-i(2E\tau_L-\phi_{\bar n}+\pi/2)],
\end{split}
\end{equation}
where $\theta=\phi_{\bar n}-\phi_n$ is the decoherence phase and $\rho\equiv\rho_{\bar n}$.
Now the answer for $\psi(N\tau_L-)$ seems evident:

\begin{equation}
\psi_{\bar n}(N\tau_L-)=\epsilon\tau_L\frac{1-\rho^Ne^{iN\theta}}{1-\rho e^{i\theta}}\exp[-i(NE\tau_L-\phi_n+\pi/2)].
\end{equation}
This conjecture is easy to verify by mathematical induction. For $t=(2\tau_L-)$ the result has been derived explicitly --- see (55). Evolving (56)
through one reflection at $t=N\tau_L$ and free propagation from $t=(N\tau_L+)$ to $t=[(N+1)\tau_L-]$ we arrive at (56) with $(N+1)$ instead
of $N$. This completes the proof.

Thus the admixture of $\bar n$ before the $N^\mathrm{th}$ collision, i.e.\ at $t=N\tau_L-$ is

\begin{equation}
|\psi_{\bar n}(N\tau_L-)|^2=\epsilon^2 \tau_L^2\frac{1+\rho^{2N}-2\rho^N\cos N\theta}{1+\rho^2-2\rho\cos\theta}
\end{equation}
The annihilation probability at the $j^\mathrm{th}$ collision is

\begin{equation}
P_a(j)=(1-\rho^2)|\psi_{\bar n}(j\tau_L-)|^2.
\end{equation}
Hence the total annihilation probability after $N$ collisions is
\begin{equation}
\begin{split}
P_a(N)=(1-\rho^2)\sum^N_{k=1}|\psi_{\bar n}(k\tau_L)|^2\\
=\frac{\epsilon^2 \tau_L^2(1-\rho^2)}{1+\rho^2-2\rho\cos\theta}\left(N+\frac{\rho^2(1-\rho^{2N})}{1-\rho^2}
-2\rho\frac{\cos\theta-\rho-\rho^N[\cos(N+1)\theta+\rho\cos N\theta]}{1+\rho^2-2\rho\cos\theta}\right).
\end{split}
\end{equation}
After several collisions the terms proportional to $\rho^N,\ \rho^{2N}$ and $\rho^{N+1}$ may be dropped since $\rho\sim 0.5$ (see (50)).
Then (59) takes the form

\begin{equation}
P_a(N)\simeq\frac{\epsilon^2 \tau_L^2}{1+\rho^2-2\rho\cos\theta}\left(N(1-\rho^2)+1-\frac{(1-\rho^2)^2}{1+\rho^2-2\rho\cos\theta}\right).
\end{equation}
Three different regimes may be inferred from (60). For very strong annihilation, i.e.\ $\rho\ll 1$

\begin{equation}
P_a(N)=\epsilon^2 \tau_L^2 N=\epsilon^2\tau_L t.
\end{equation}
For complete decoherence at each collision, i.e.\ for $\theta=\pi$

\begin{equation}
P_a(N)=\epsilon^2 \tau_L^2\left(N\frac{1-\rho}{1+\rho}+\frac{\rho(2-\rho)}{(1+\rho)^2}\right)\simeq\frac{1-\rho}{1+\rho}\epsilon^2\tau_L t.
\end{equation}
For the (unrealistic) situation when $\theta=0$

\begin{equation}
P_a(N)=\epsilon^2 \tau_L^2\left(N\frac{1+\rho}{1-\rho}-\frac{\rho(2+\rho)}{(1-\rho)^2}\right)\simeq\frac{1+\rho}{1-\rho}\epsilon^2\tau_L t.
\end{equation}
For the values of $\rho$ and $\theta$ corresponding to the optical potentials (17)--(18) the quantity $Q_a(N)=(\epsilon^2\tau_L^2 N)^{-1}=
(\epsilon^2\tau_Lt)^{-1}P_a(N)$ calculated according the exact equation (59) is displayed in Fig.~\ref{Fig1}. This figure shows that the linear
time dependence settles after about 10 collisions with the trap walls. The asymptotic value of $Q_a(N)$ which may be called the enhancement
factor is $1.5\div 2$ depending on the wall material.

\begin{figure}[h]
\epsfbox{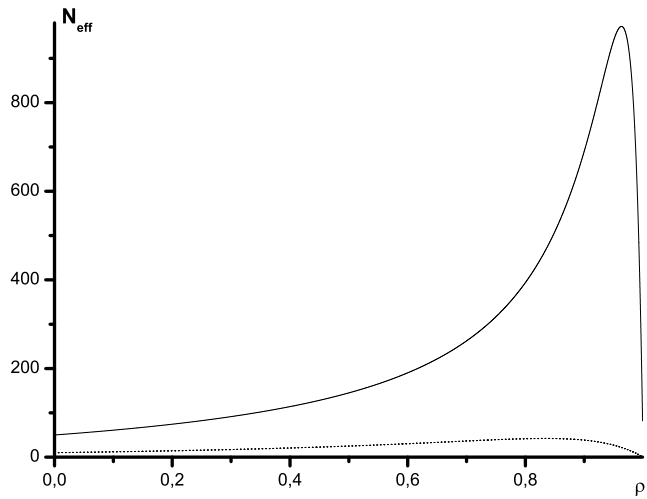}
\caption{Plot of $N_{eff}$ versus $\rho$ dependence. Solid line is for the number of collisions $N=50$,
dashed one corresponds to $N=10$.}
\label{Fig2}
\end{figure}

Proposals have been discussed in the literature \cite{Golub, PRIgn} to compensate the decoherence phase $\theta$ by applying the external
magnetic field. Suppose that in such a way the regime $\theta=0$ may be achieved. Also suppose that one can vary the reflection coefficient
$\rho$ in a whole range by varying the trap material. For such an ideal situation we plot in Fig.~\ref{Fig2} the quantity $N_{eff}(\rho)$
defined as

\begin{equation}
P_a(N)=\epsilon^2\tau_L^2N_{eff}(\rho).
\end{equation}
Thus defined $N_{eff}(\rho)$ obviously depends also on the number of collisions $N$ and in Fig.~\ref{Fig2} the results for $N=10$ and $N=50$
are presented. This figure shows what can be expected from the trap experiments in the most favorable though hardly realistic scenario.

\section{Concluding Remarks}

We have reexamined the problem of $n-\bar n$ oscillations for UCN in a trap. Our aim was to present a clear formulation
of the problem, to calculate the amplitude of the $\bar n$ component for arbitrary observation time and for any given
reflection properties of the trap walls. We have shown that for physically relevant observation time (i.e.\ for the time interval less
than $\beta$-decay time) the process of $n-\bar n$ oscillations is described in terms of the wave packets while the standing waves
regime may settle only at later times. By calculating the difference between $n$ and $\bar n$ collision times the new light
has been shed on the decoherence phenomena. For the first time an exact equation has been derived for the annihilation probability
for an arbitrary number of collisions with the trap walls. In line with the conclusions of the previous authors on the subject this probability
grows linearly with time. We have calculated the enhancement factor entering into this linear time dependence and found this
factor to be $1.5\div 2$ depending on the reflection properties of the wall material.

Despite the extensive investigations reviewed in this article and the results of the present paper the list of problems for further work
is large. The central and most difficult task is to obtain reliable parameters of the optical potential for antineutrons. The beam of $\bar n$
with energy in the range of $10^{-7}\ \mathrm{eV}$ will be hardly accessible in the near future. Therefore work has to be continued along
the two lines mentioned above --- to deduce the parameters of the optical potential from the level shifts in antiprotonic atoms and to
construct reliable optical models which can be confronted with the available experimental data on $\bar n$-nuclear interaction at higher
energies. In the forthcoming publication we plan to present numerical calculation of the time evolution of the wave packet into standing
waves as well as to discuss some features of $n-\bar n$ oscillations in the eigenfunctions basis which were not discussed in Ref.~\cite{Marsh}.
Another task is to perform calculation for the concrete geometry of the trap an realistic spectrum of the neutron beam. To this end one needs
an input corresponding to a concrete experimental setting.

The authors are grateful to Yu.~A.~Kamyshkov for interest to the work and discussions. Useful remarks and
suggestions from L.~N.~Bogdanova, F.~S.~Dzheparov, A.~I.~Frank, 
V.~D.~Mur, V.~S.~Popov, G.~V.~Danilian are gratefully acknowledged.

One of the authors (B.~O.~K.) expresses his gratitude for financial support to V.~A.~Novikov, L.~B.~Okun, INTAS grant 110
and K.~A.~Ter-Martirosian scientific school grant NSchool--1774.

\end{document}